%% file: main.tex
\documentclass[sigconf]{acmart}
\AtBeginDocument{%
  \providecommand\BibTeX{{%
    \normalfont B\kern-0.5em{\scshape i\kern-0.25em b}\kern-0.8em\TeX}}}

\copyrightyear{2023}  \acmYear{2023}  \setcopyright{rightsretained}  \acmConference[CHI EA '23]{Extended Abstracts of the 2023 CHI Conference on Human Factors in Computing Systems}{April 23--28, 2023}{Hamburg, Germany} \acmBooktitle{Extended Abstracts of the 2023 CHI Conference on Human Factors in Computing Systems (CHI EA '23), April 23--28, 2023, Hamburg, Germany}\acmDOI{10.1145/3544549.3585861} \acmISBN{978-1-4503-9422-2/23/04}

\begin{document}
\newcommand{\tool}{\textit{PonziLens}}
\title{Code Will Tell: Visual Identification of Ponzi Schemes on Ethereum}


\author{Xiaolin Wen}
\orcid{1234-5678-9012}
\affiliation{%
  \institution{Sichuan University}
  \streetaddress{P.O. Box 1212}
  \city{Chengdu}
  \country{China}
}
\affiliation{
  \institution{Singapore Management University}
  \country{Singapore}
}
\email{wenxiaolin@stu.scu.edu.cn}

\author{Kim Siang Yeo}
\affiliation{%
  \institution{Singapore Management University}
  \country{Singapore}}
\email{ks.yeo.2021@mitb.smu.edu.sg}

\author{Yong Wang}
\authornote{The corresponding author.}
\affiliation{%
  \institution{Singapore Management University}
  \country{Singapore}}
\email{yongwang@smu.edu.sg}

\author{Ling Cheng}
\affiliation{%
  \institution{Singapore Management University}
  \country{Singapore}}
\email{lingcheng.2020@phdcs.smu.edu.sg}

\author{Feida Zhu}
\affiliation{%
  \institution{Singapore Management University}
  \country{Singapore}}
\email{fdzhu@smu.edu.sg}

\author{Min Zhu}
\affiliation{%
  \institution{Sichuan University}
  \city{Chengdu}
  \country{China}}
\email{zhumin@scu.edu.cn}
\renewcommand{\shortauthors}{Xiaolin Wen, Kim Siang Yeo, Yong Wang, Ling Cheng, Feida Zhu, and Min Zhu.}

\begin{abstract}
Ethereum has become a popular blockchain with smart contracts for investors nowadays.
Due to the decentralization and anonymity of Ethereum, Ponzi schemes have been easily deployed and caused significant losses to investors.
However, there are still no explainable and effective methods to help investors easily identify Ponzi schemes and validate whether a smart contract is actually a Ponzi scheme.
To fill the research gap, we propose \tool{}, a novel visualization approach to help investors achieve early identification of Ponzi schemes by investigating the operation codes of smart contracts.
Specifically, we conduct symbolic execution of opcode and extract the control flow for investing and rewarding with critical opcode instructions.
Then, an intuitive directed-graph based visualization is proposed to display the investing and rewarding flows and the crucial execution paths, enabling easy identification of Ponzi schemes on Ethereum.
Two usage scenarios involving both Ponzi and non-Ponzi schemes demonstrate the effectiveness of \tool{}.
\end{abstract}

\ccsdesc[500]{Human-centered computing~Visual analytics}
\ccsdesc[500]{Human-centered computing~Information visualization}

\keywords{Ponzi scheme, visual identification, Ethereum, visual analytics}

\begin{teaserfigure}
  \includegraphics[width=\textwidth]{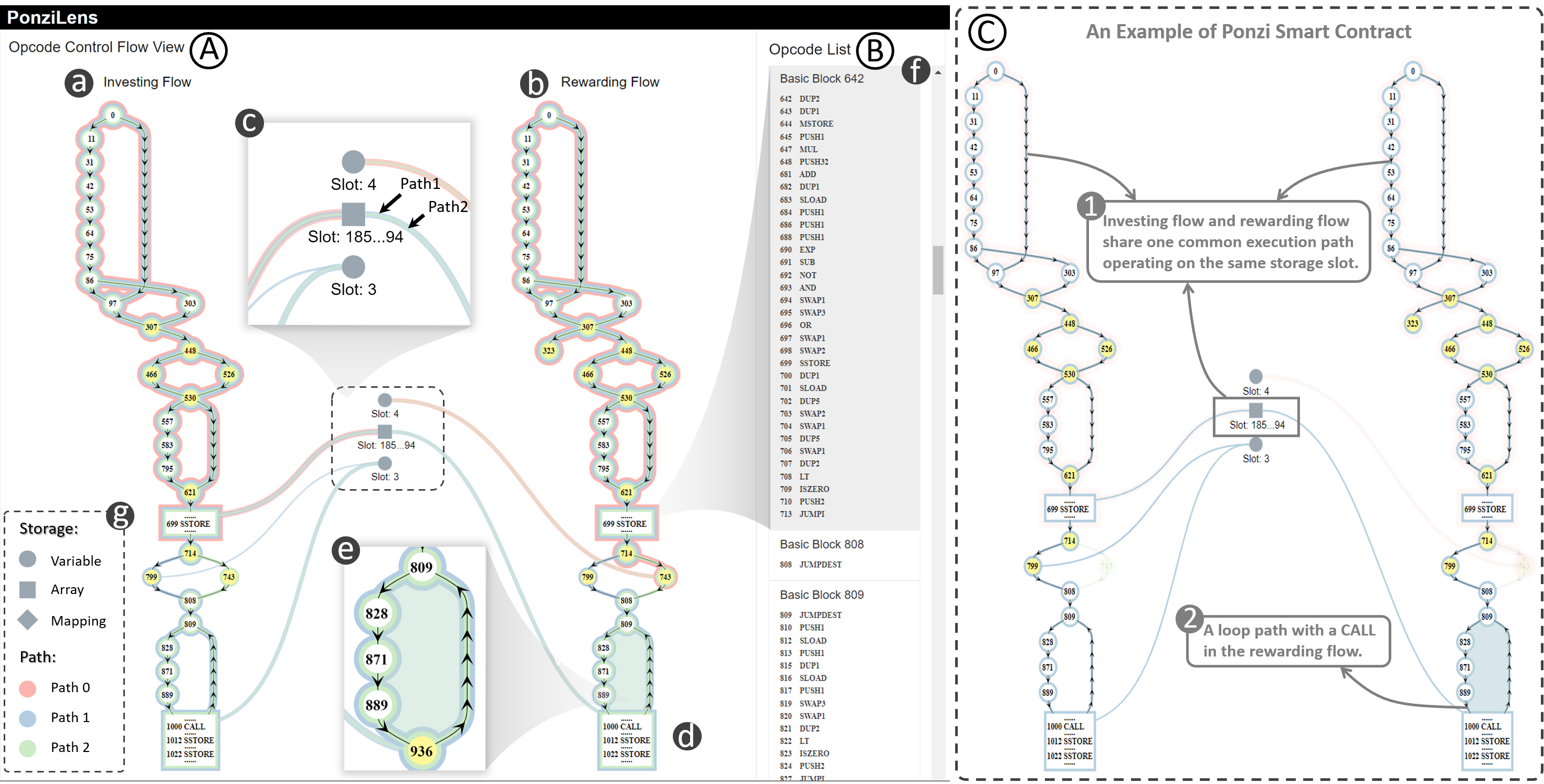}
  \caption{The user interface of \tool{} consists of (A) \textit{Opcode Control Flow View} and (B) \textit{Opcode List}.
  (A) The \textit{Opcode Control Flow View} shows (a) investing flow, (b) rewarding flow, and (c) storage interactions — all of which are critical for identifying a Ponzi smart contract. (B) The \textit{Opcode List} shows all the original operation codes of a smart contract, where the operation code of a basic block can be highlighted (f). (d) shows that a basic code block in control flow can be unfolded to check the critical instructions within it. (e) shows an execution loop within a CALL instruction. (g) is the legend illustrating the storage type and the aggregated paths. (C) shows the Opcode Control Flow View with the aggregated path in blue \textit{Path1} highlighted.}
  \label{fig:teaser}
  \Description{Figure 1. fully described in the text.}
\end{teaserfigure}


\maketitle

\input{src/1_intro}

\input{src/2_relatedwork}

\input{src/3_method}

\input{src/4_evaluation}
\input{src/5_conclusion}
\begin{acks}
This work was done during Xiaolin Wen's internship at the Singapore Management University (SMU) under the supervision of Dr. Yong Wang. This work was supported by the Singapore Ministry of Education (MOE) Academic Research Fund (AcRF) Tier 1 grant (Grant number:
20-C220-SMU-011), Lee Kong Chian Fellowship awarded to Dr. Yong Wang by SMU, and the National Natural Science Foundation of China (Grant number: 62172289).
We would like to thank the anonymous reviewers for their
feedback.
\end{acks}

\bibliographystyle{ACM-Reference-Format}
\bibliography{ref}










\end{document}

%% file: src/1_intro.tex
\section{Introduction}





With the prevalence of blockchain, Ethereum, a blockchain-based system, has become an increasingly popular way for investors to carry out decentralized, secure and anonymous
transactions
without an intermediate third party's credit endorsement~\cite{dannen2017introducing,wood2014ethereum}.
Ethereum incorporates smart contracts that define the transaction rules in the form of source code
on the blockchain. Such smart contracts will be executed automatically on the Ethereum Virtual Machine (EVM) once the predefined conditions in the contract are met~\cite{buterin2014next,szabo1996smart}.
All the transaction records and smart contract codes are publicly available and immutable on Ethereum.

Unfortunately, scammers have also leveraged the anonymity and immutability of blockchain and deployed various ``trustworthy'' frauds on Ethereum to
cheat investors of their money, or Ether — the scarce digital money on Ethereum.
Among all the frauds, Ponzi schemes~\cite{artzrouni2009mathematics,moore2012postmodern} are a popular investment scam on Ethereum that lure investors with a promise of high profits that are actually from the invested Ether of subsequent new investors, instead of actual investment appreciation income. 
Since all the byte codes of smart contracts on Ethereum are publicly available, it gives investors an illusion that smart contracts on Ethereum are credible, making them tend to trust the smart contracts.
Also, due to the anonymity and immutability of blockchain, it is difficult to track and identify the fraudsters and also unable to revoke the Ponzi scheme transactions once the transactions are written into the blockchain.
All these factors have made Ponzi schemes easy to be deployed on Ethereum and Ponzi schemes have caused significant economic losses to investors on Ethereum.
According to
Chen et al.~\cite{chen2021sadponzi}, Ponzi schemes on Ethereum have led to losses of more than US\$17 million by 2021.

Early studies have first leveraged transaction data for money flow analysis, and further detected Ponzi schemes on blockchain~\cite{toyoda2017identification,bartoletti2018data,toyoda2019novel}. They intrinsically require that at least a group of investors have fallen into the trap of Ponzi schemes and cannot work for early detection of Ponzi schemes.
More recent Ponzi scheme detection techniques~\cite{fan2021spsd,chen2021sadponzi,sun2020early,zhang2021code,hu2021scsguard,he2022ctrf} have further investigated the operation code (opcode) of smart contracts on Ethereum.
By considering the characteristics of opcodes (e.g., the operator frequency) of Ponzi schemes, they can achieve an early identification of Ponzi schemes before any investors are trapped by a Ponzi scheme.
However, such techniques rely on holistic features of opcodes such as the operator frequency and cannot adapt to various Ponzi schemes to achieve a consistently high detection accuracy.
Also, they totally ignore the semantic meaning of opcodes~\cite{chen2021sadponzi}, making it difficult for investors to understand why a smart contract is predicted as a Ponzi scheme. 
An explainable and effective way to help investors identify Ponzi schemes on Ethereum is still missing.

In this paper, we fill the research gap by informing investors of the semantic meaning of the opcodes of smart contracts to facilitate Ponzi scheme identification on Ethereum.
Specifically, we propose \tool{}, a visualization approach to show the investing and rewarding flows of smart contracts as well as their relations, revealing the essential characteristic of Ponzi schemes, i.e., \textit{whether the reward of prior investors directly comes from the investments of subsequent new investors}.
However, it is a challenging task due to the vast differences in the opcodes of various smart contracts as well as the difficulty of making common investors easily understand the function of complex opcodes.
Inspired by prior studies~\cite{krupp2018teether,contro2021ethersolve}, we extract the Control Flow Graph from the opcode of a smart contract on Ethereum via symbolic execution, and further identify execution paths relevant to the investing and rewarding process of smart contracts by using crucial opcode instructions like \textit{CALL, CALLER, SSTORE} and \textit{SLOAD}. Some crucial features indicating a Ponzi scheme, including storage stacks shared by investing and rewarding flows and opcode loops, are also extracted.
Then, we propose an intuitive directed-graph based visualization to show the investing and rewarding flows of smart contracts, where the crucial execution paths and common storage are also explicitly highlighted.
\tool{} clearly visualizes all the Ether flows within a smart contract, and helps investors easily identify Ponzi schemes on Ethereum.
To the best of our knowledge, it is the first time that the semantically-meaningful Ether flows of smart contracts have been visualized for Ponzi scheme identification.
We showcase two usage scenarios, where both a Ponzi scheme and non-Ponzi scheme are investigated, to demonstrate the usefulness of \tool{}. In summary, the contributions of this paper can be summarized as follows:
\begin{itemize}
    \item A novel visualization approach, \tool{}, to inform investors of the investing and rewarding flows of a smart contract and facilitate easy identification of Ponzi schemes on Ethereum.
    
    \item Two usage scenarios involving both Ponzi and non-Ponzi schemes to demonstrate the usefulness of our approach.
\end{itemize}

%% file: src/2_relatedwork.tex
\section{Related Work}
This paper is related to prior research on \textit{automated detection of Ponzi schemes} and \textit{visual analytics for blockchain data}. 

\textbf{Automated detection of Ponzi schemes}
Many approaches have been developed to achieve automated detection of Ponzi schemes in blockchain by analyzing the transaction data and the source code of smart contracts.
For transaction-based approaches, they often leverage machine learning techniques, such as ordered boosting~\cite{fan2021spsd}, attention neural networks~\cite{hu2021scsguard} and behaviour forest~\cite{sun2020early}, to learn the characteristics of Ponzi schemes and achieve automated Ponzi detection.
These approaches intrinsically cannot work for early detection of Ponzi schemes before any transactions of a Ponzi scheme are invoked.
For smart contract-based approaches, they mainly attempt to extract distinctive features of Ponzi schemes from smart contracts via different ways such as symbolic execution of opcode~\cite{chen2021sadponzi} and code attribution for de-anonymizing smart contracts~\cite{linoy2021anonymizing}.
Also, some recent works have combined transaction data with the source code of smart contract together for automated detection of Ponzi schemes~\cite{jung2019data, chen2019exploiting,
zhang2021code,
he2022ctrf}.
However, almost all the smart contract-based approaches rely on extracting high-level features and ignore the semantic meanings of the opcodes of smart contracts. It makes their result difficult to be understood by investors, which will be addressed in our approach.


\textbf{Visual analytics for anomaly detection on blockchain}
Prior studies have developed visual analytics approaches to facilitate anomaly detection on blockchain.
Transaction data capture all the interactions between entities, and has been explored for helping investors identify different anomalies 
on blockchain (especially bitcoin blockchain) in a general way.
For example, Blockchain explorer~\cite{7945049}, Biva~\cite{oggier2018biva}, BitVis~\cite{sun2019bitvis} and Bitconeview~\cite{7312773} provided detailed statistics of transactions on bitcoin blockchain, and also visualized the relations between different wallet addresses and transactions.
These approaches can effectively reveal some anomalies such as money laundering and Ponzi schemes, as these anomalies show obvious characteristics in their transactions.
Other visual analytics methods have also investigated the transaction data and highlighted some transaction features that are specific for one specific anomaly. For instance, Ahmed et al.~\cite{ahmed2018tendrils} used taint tracking to trace the trail of stolen money back to its owner. Balthasar et al.~\cite{balthasar2017analysis} visualized the unique transaction patterns of money laundering such as the mixing of bitcoins, facilitating money laundering identification.
Wen et al.~\cite{wen2023nft} proposed NFTDisk, a visual analytic system, to help investors detect wash trading in NFT markets.
Also, a few visualization approaches have also been proposed to analyze the source code of smart contracts for different purposes such as facilitating smart contract development~\cite{tan2020latte} and Solidity code representation~\cite{pierro2021smart}.
However, none of the above studies has attempted to visualize the source code of smart contracts for Ponzi scheme detection on Ethereum, which is the focus of this paper.

%% file: src/3_method.tex
\section{Background} \label{Sec:background}

This section introduces the overall background of Ethereum and Ponzi schemes.


\textbf{Data on Ethereum.}
The EVM is a \textit{stack}-based architecture, where \textit{stack} is an internal place to store temporal variables~\cite{wood2014ethereum}.
Besides \textit{stack}, the EVM also stores data in two other places~\cite{wood2014ethereum,chen2021sadponzi}: \textit{memory} and \textit{storage}.
\textit{Memory} is a byte array
storing the
data for function execution and \textit{storage} is used to permanently keep data 
on
the Ethereum blockchain. Since Ponzi schemes need to return Ether to past investors, the information of all investors is stored in \textit{storage}.
\textbf{Opcodes of smart contracts.}
For smart contracts to work, they have to be compiled from their high-level languages (e.g., Solidity) into opcodes (also called operation codes) that can be
executed by the EVM~\cite{wood2014ethereum}. According to prior research~\cite{chen2021sadponzi} and our own observations, Ponzi smart contracts use four critical opcode instructions: \textit{CALLER}, \textit{CALL}, \textit{SSTORE}, and \textit{SLOAD}.
For a Ponzi smart contract,
\textit{CALLER} adds the address of the account that used a smart contract to the \textit{stack}, and is used to retrieve the new investor's account address. \textit{SSTORE} copies data from the \textit{stack} to \textit{storage} and is used to permanently store the information of new investors. \textit{SLOAD} reads a value from the \textit{storage}, and retrieves information of previous investors in a Ponzi smart contract. \textit{CALL} is used to transfer Ether to an address. Ponzi smart contracts use the instruction \textit{CALL} to transfer Ether to the previous investor account addresses obtained using \textit{SLOAD}.

\textbf{Basic blocks, execution paths and control flow graph.}
Opcodes can be separated into groups of basic opcode blocks, where each one affects the \textit{stack} in the same way and ends with either a condition leading to another basic opcode block or a termination instruction. Depending on the design logic of a smart contract, there can be different execution sequences of these basic opcode blocks, which are called \textit{execution paths} in this paper. Each execution path carries out a task specified in the smart contract. We use a Control Flow Graph (CFG) to represent all the possible execution paths.

\textbf{Ponzi schemes at opcode level.}
According to SADPonzi~\cite{chen2021sadponzi},
the smart contracts of Ponzi schemes
involve two types of critical actions that can be captured through opcodes: \textit{investing} and \textit{rewarding}.
For investing actions, an investor invokes a transaction of a smart contract, which requires saving the investor information to the \textit{storage} for future payment of rewards. 
Thus, an execution path that leverages \textit{SSTORE} to save the data recorded by \textit{CALLER} to the \textit{storage} is considered as investing.
The \textit{rewarding} actions of Ponzi schemes refer to paying the profits from each new investment to prior investors, which must use the instruction \textit{CALL}.
Also, the \textit{storing} actions are critical for identifying Ponzi schemes, as it is necessary to check whether the location storing the investor information in an \textit{investing} action is the same as the storage location recording the investor information in the subsequent rewarding action.

\section{Our Method}
\begin{figure}[htb]
  \centering
       \setlength{\abovecaptionskip}{0.1cm}
    \setlength{\belowcaptionskip}{0.cm}
  \includegraphics[width=0.8\linewidth]{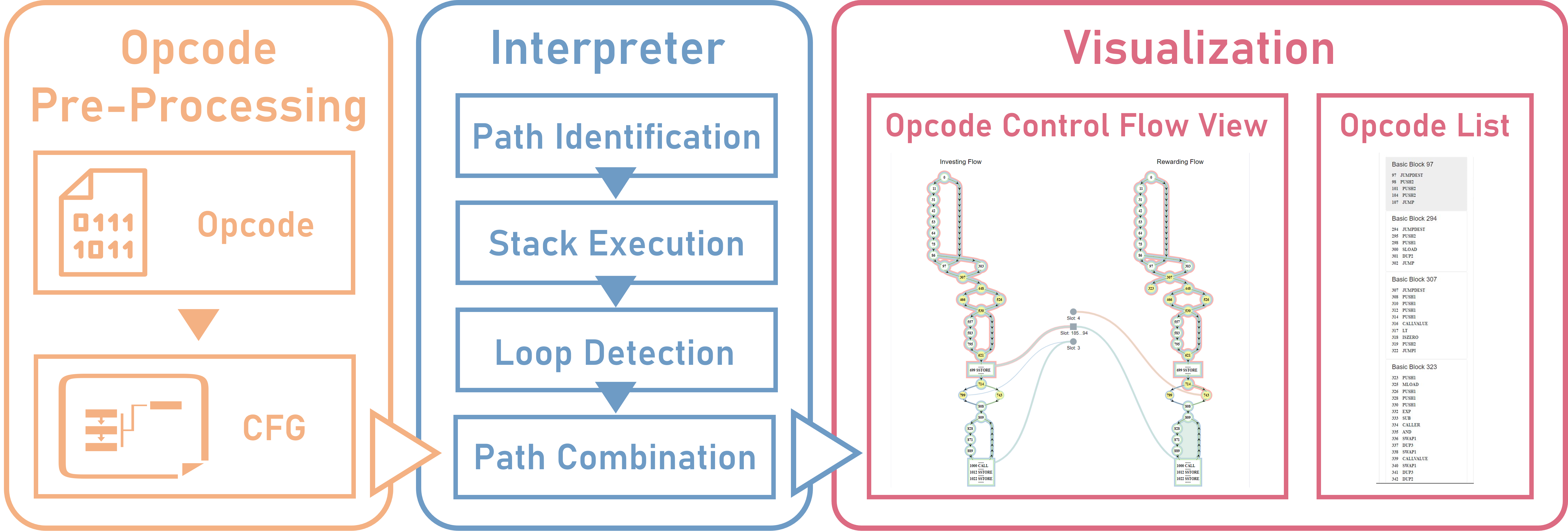}
  \caption{\label{system}
        The architecture of \tool{} consists of three modules: opcode pre-processing, interpreter, and visualization.
           }
\Description{Figure 2. fully described in the text.}
\end{figure}

\tool{} consists of three modules (Fig.~\ref{system}): \textit{opcode pre-processing}, \textit{interpreter} and \textit{visualization}.
In the opcode pre-processing module, we collect the opcodes of a specific smart contract from EtherScan\footnote{\url{https://etherscan.io/}}, a Block Explorer for Ethereum, and input them into teEther~\cite{krupp2018teether} to identify all basic opcode blocks and create the corresponding CFG. The interpreter module uses these basic opcode blocks and CFG to identify the crucial investing and rewarding execution paths, and the \textit{storage} slots used by these paths. The visualization module shows the possible investing and rewarding paths and their interactions with the \textit{storage} slots, enabling investors easily identify the possible Ponzi schemes in an intuitive way. Lastly, an opcode list is also included for a detailed investigation of the opcodes of a smart contract.


\subsection{Interpreter}
The interpreter is designed to extract the critical information from the opcodes and CFG, and consists of four steps executed sequentially.

\textbf{Path identification for a smart contract}
Path identification aims to collect all potential investing and rewarding paths from the CFG.
We build all execution paths as Directed Acyclic Graphs (DAG), which are further categorised into investing paths that contain the opcode instructions \textit{CALLER} and \textit{SSTORE} and rewarding paths containing the opcode instructions \textit{CALL}. 


\textbf{Stack execution}
Next, we need to determine if the investing and rewarding paths are properly executed by the EVM through the stack.  In a Ponzi smart contract, an investing path should present a \textit{CALLER} instruction on its \textit{stack} when executing the instruction \textit{SSTORE}, indicating that the investor's information is stored in the \textit{Storage}. A rewarding flow should pick up an address obtained with the instruction \textit{SLOAD} when transferring Ether to the previous investors using the instruction \textit{CALL}. Through a process called stack execution, \textit{teEther} ~\cite{krupp2018teether} uses z3 ~\cite{moura2008z3} to run an execution path and identify what it does to the \textit{stack}. We apply this method to check all the identified paths, and keep the ones that exhibit investing and rewarding behaviours on the \textit{stack}. Stack execution also provides us with information on the \textit{storage} slots involved in either the investing or rewarding paths, or both. Each slot is a number when used to store state variables, and SHA-256 hash is used to retrieve data structures such as an array or mappings on \textit{storage}. This slot information will be used in the visualization module. 



\textbf{Loop detection.}
Ponzi schemes often use a rewarding execution path containing a loop starting with the instruction \textit{CALL} to reward multiple investors. We modified the method used during path identification to determine which rewarding path exhibits this characteristic.



\textbf{Path combination}
We combined investing and rewarding paths performed exactly the same way during stack execution into aggregated paths with a set of edges and nodes from the original execution paths to maximize the efficiency of our analysis. 

\subsection{Visualization}\label{Sec:visualization}
The visualization module helps investors to explore the features of a smart contract interactively and validate whether it is a Ponzi scheme.
Specifically, an Opcode Control Flow View (Fig.~\ref{fig:teaser}A) shows the potential execution paths of a smart contract when an investor invokes an investment transaction using a smart contract. 
An opcode List (Fig.~\ref{fig:teaser}B) is incorporated to show the details  of each basic opcode block in the CFG, helping investors further verify the insights from the Opcode Control Flow View.

Ponzi schemes have distinctive characteristics in their opcodes (Sec.\ref{Sec:background}), which has also guided our visualization design. Specifically, we have considered the following two major opcode characteristics of Ponzi schemes (Fig.~\ref{fig:teaser}C):

\textbf{C1. Investing flow and rewarding flow share one common execution path operating on the same storage slot.}
For a Ponzi scheme, the target address of the Ether transferring in the rewarding flow is obtained from the same \textit{storage} slot where the prior investors' information is stored during the investing flow, as the invested Ether by a new investor will be paid as rewarding to prior investors.
Also, such a payment is directly done when a new investment transaction is provoked, making the investing flow and rewarding flow share one common execution path.

\textbf{C2. A loop path with a \textit{CALL} instruction in the rewarding flow.}
The rewarding flow of the smart contract of Ponzi schemes often needs to pay Ether to multiple prior investors, which results in a loop in the execution path. Since \textit{CALL} is necessary for transferring Ether in opcode of Ethereum, the loop also contains \textit{CALL}.





\subsubsection{Opcode Control Flow View}
The Opcode Control Flow View (Fig.~\ref{fig:teaser}A) is designed to show the potential investing and rewarding paths and their interactions with the \textit{storage} in EVM.
The Ponzi Detection View consists of three parts: investing control flow (abbreviated as investing flow in Fig.~\ref{fig:teaser}a), rewarding control flow (abbreviated as rewarding flow in Fig.~\ref{fig:teaser}b), and storage interactions (Fig.~\ref{fig:teaser}c). 

\textbf{Investing flow and rewarding flow}
Investing control flow provides an overview of all aggregated investing paths. 
As shown in Fig.~\ref{fig:teaser}a, the investing flow is a directed graph. The basic opcode blocks in the CFG are represented by circular nodes with a label indicating its block index, and the execution path order is indicated by a black curve with arrows. As there can be overlaps between aggregated paths from the interpreter, we use a different color to wrap around each path's associated basic opcode blocks to represent it. This way, the investor can quickly identify the basic blocks that each path passes through. Similarly, the rewarding control flow shows all aggregated rewarding paths, which uses the same visual encodings as the investing flow.
The investing flow and rewarding flow in one execution path are encoded in the same color to show that they will be executed in the same contract call. Given that only code loops with the \textit{CALL} instruction are important, we visually encode them with the same color used to encode the rewarding paths they originate from. To help the investor visually differentiate between the directed-graph and the code loops, we also fill in the area that the loop path encloses with the same color. 
Furthermore, the basic blocks containing  the critical opcodes like CALLER, CALL, and SSTORE, are indicated with a yellow background (Fig.~\ref{fig:teaser}A and Fig.~\ref{fig:teaser}B) can be unfolded by clicking  to show the critical opcodes in it (Fig.~\ref{fig:teaser}d).
The investors are also allowed to highlight one of the paths in the Opcode Control Flow View by hovering above the path.

\textbf{Storage interactions}
The storage interactions are designed to display how the investing flow and rewarding flows interact with the \textit{storage} in EVM and help investors verify whether the new investment is directly transferred to previous investors.
We draw all \textit{storage} slots accessed by the investing and rewarding flows, as shown in Fig.~\ref{fig:teaser}c.
The type of data in each \textit{storage} slot is encoded by different glyphs. In the Opcode Control View, the circle is used to visually represent a state variable, and the rectangle is used to represent an array. (Fig.~\ref{fig:teaser}g).
A line is used to intuitively represent interactions between a path and one or more \textit{storage} slots (Fig.~\ref{fig:teaser}c). When an investing path interacts with a \textit{storage} slot, this visualizes the storing of investor information to that \textit{storage} slot. When a rewarding path interacts with a \textit{storage} slot, this visualizes the retrieval of past investor information from that \textit{storage} slot so that Ether can be transferred to them. Interactions that belong to the same execution path are encoded with the same color. 
By analyzing the storage interactions of the investing flow and rewarding flow, it is easy and intuitive to verify whether the contract transfers Ether to the previous investors.

\subsubsection{Opcode List}
To help investors further confirm the insights obtained from the Opcode Control Flow View, we also show the original opcodes list separated according to the basic opcode block index.
All nodes unfolded by investors in the Opcode Control Flow View are highlighted in the Opcode List, and the Opcode List will scroll to the position of the last block clicked by the investor.




%% file: src/4_evaluation.tex
\section{Usage Scenario}
We showcase two usage scenarios with both Ponzi and non-Ponzi smart contracts to demonstrate the effectiveness of \tool{}. 


\subsection{Scenario 1: A Ponzi Smart Contract}
We use \tool{} to explore the smart contract of a confirmed Ponzi scheme\footnote{Address: 0x0b230b071008bbb145b5bff27db01c9248f486b9} that has been used by prior studies~\cite{bartoletti2020dissecting,chen2021sadponzi}.

Fig.~\ref{fig:teaser}A provides an overview of the investing and rewarding flows of this smart contract,
where some opcode blocks in these paths lead to the storage slots as shown in Fig.~\ref{fig:teaser}c.
We can see that the paths are encoded in three colors (red, blue, and green), indicating
that there are three aggregated paths (i.e., \textit{Path0}, \textit{Path1}, and \textit{Path2}) in this smart contract.
Figs.~\ref{fig:teaser}a and \ref{fig:teaser}b show that all the three paths are involved in both investing and rewarding flows, 
while Fig.~\ref{fig:teaser}c shows that \textit{Path1} and \textit{Path2}, indicated by the blue and green colors, congruently link their investing and rewarding flows through the same \textit{storage} slot \textit{185...94}.
The insights here are two-fold.
First, both investing and rewarding will be triggered when a transaction of this smart contract is provoked.
Secondly, the invested Ether by new investors is probably paid to prior investors as rewards, as \textit{Path1} and \textit{Path2}, involved in both the investing flow and rewarding flow, operated on the same \textit{storage} slot \textit{185...94}.



To clearly check \textit{Path1}, we can
hover above the blue path to highlight it,
as shown in Fig.~\ref{fig:teaser}C.
Also, we can see that there is an execution loop in the rewarding flow with the \textit{CALL} instruction presented (Fig.~\ref{fig:teaser}e), indicating that
the smart contract is recursively sending Ether to
the addresses of multiple prior investors.


By comparing all the above observations with the two major characteristics of a Ponzi scheme (Sec.~\ref{Sec:visualization}), it is safe to confirm that this smart contract is a Ponzi scheme.

\subsection{Scenario 2: A Smart Contract for Charity}

We further use \tool{} to explore a smart contract for charity, since they are not Ponzi schemes but also transfer Ether from a large number of donors to the address of a charity,
Specifically,
we explored is EthPledge\footnote{Address: 0x10Ec03b714A2660581040c1A0329d88e381cA603}\footnote{\url{https://www.ethpledge.com/}}, a decentralized smart contract on Ethereum that allows people to donate money to a charity.

\begin{figure}[htb]
  \centering
  \includegraphics[width=0.7\linewidth]{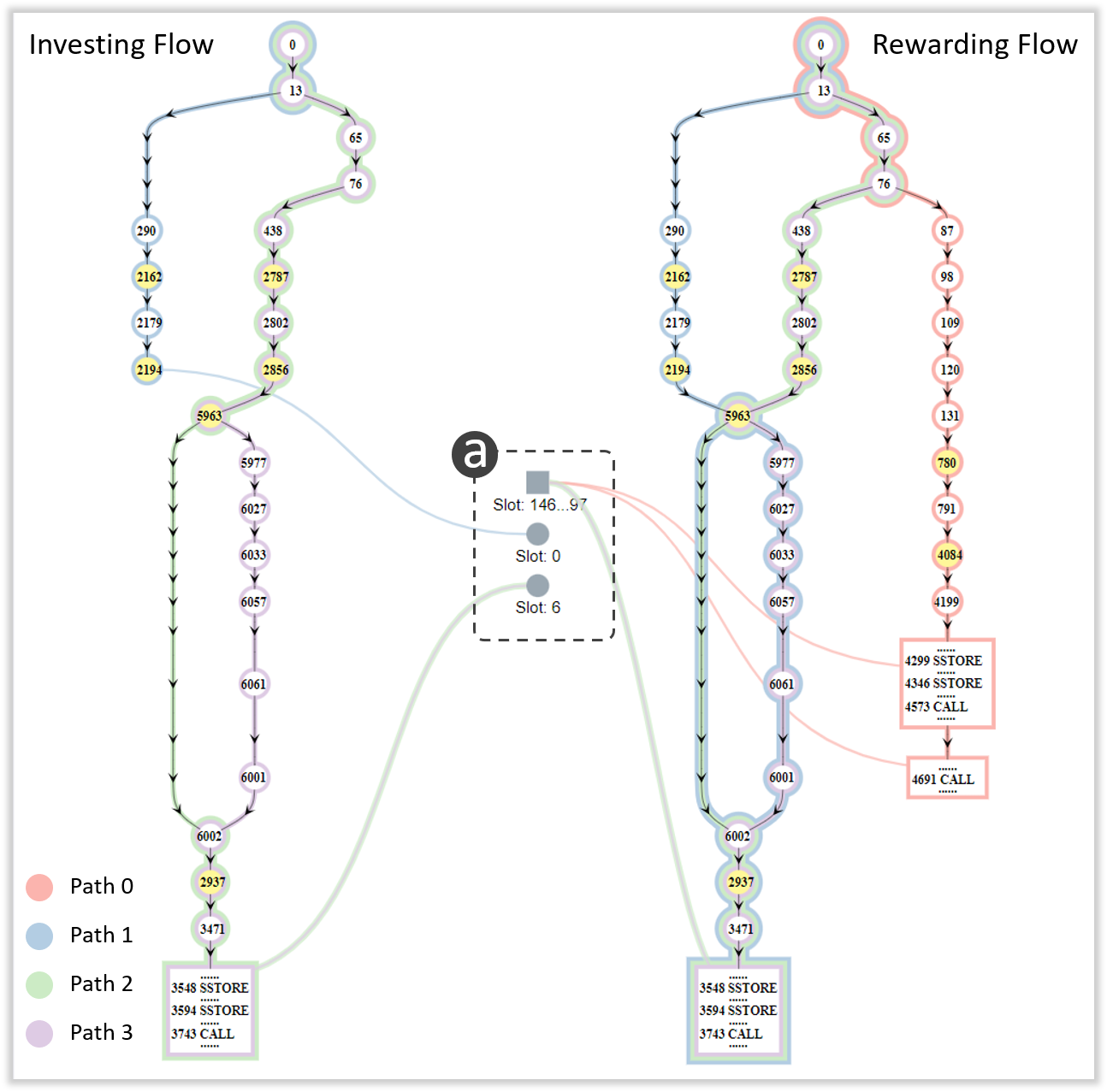}
  \caption{\label{case2}
        The opcode control flow of \textit{EthPledge}, a smart contract for charity.
        \tool{} shows the investing flow and the rewarding flow, as well as their interactions with storage slots (a).
           }
    \Description{Figure 3. fully described in the text.}
\end{figure}

Fig.~\ref{case2} shows the investing and rewarding flows of this smart contract. It demonstrates that this contract receives and transfers Ether, which is natural for a charity.
However, there are no congruent execution paths from the investing flow to the rewarding flow linked by a location in \textit{storage} (Fig.~\ref{case2}a). 
We can see that the execution paths involved in the investing and rewarding flows use different \textit{storage} slots (Fig.~\ref{case2}a), indicating that investments cannot be transferred to prior investors.
Given that the key characteristics of Ponzi schemes mentioned in Sec.~\ref{Sec:visualization} is not seen here,
we can confidently conclude that this smart contract is NOT a Ponzi scheme.

Also, it is interesting to see that all the interactions between \textit{storage} and the rewarding flow
occur at the same \textit{storage} slot \textit{146...97} (the rectangle of Fig.~\ref{case2}a). It indicates that Ether is always transferred back to the addresses stored in an array of \textit{storage}. This array probably records the information of all the charity organizations.

%% file: src/5_conclusion.tex
\section{Discussion and Conclusion}
In this paper, we propose a novel visual analytics system, \tool{}, for early identification of Ponzi smart contract.
\tool{} can visualize all possible investing or rewarding execution paths in an aggregated manner as well as their interactions with the EVM storage, which reveals the critical features for identifying a Ponzi smart contract. 
We present two usage scenarios with two real smart contracts on Ethereum blockchain.
The results indicate that \tool{} can help investors easily identify Ponzi schemes on Ethereum.


However, \tool{} is not without limitations.
The current visual design may suffer from scalability issues when a smart contract has a huge number of basic opcode blocks and execution paths. 
In the future, we plan to improve the scalability of \tool{} by supporting the hierarchical aggregation of basic opcode blocks and execution paths.
Further, it will be interesting to explore how our approach can be extended to the early detection of other smart contract frauds like
honeypot contracts and pump-and-dump schemes.
Besides, a quantitative user study is necessary future work to further evaluate the effectiveness and usability of \tool{}.